\documentclass{emulateapj}

\usepackage{graphicx}
\shorttitle{DASCH}
\shortauthors{Laycock et al.}

\begin{document}


\title{Digital Access to a Sky Century at Harvard. II:  Initial Photometry and Astrometry} 


\author{S. Laycock, S. Tang, J. Grindlay, E. Los, R. Simcoe, D. Mink}
\affil{Harvard-Smithsonian Center for Astrophysics, 60 Garden St, Cambridge, MA, 02138}



\begin{abstract} 
Digital Access to a Sky Century @ Harvard (DASCH) is a project to digitize the collection of $\sim$500,000 glass photographic plates 
held at Harvard College Observatory. The collection spans the time period from 1880 to 1985, during which time every point on the sky has been observed approximately 500 to 1000 times.  
In this paper we describe the results of the DASCH commissioning run, during which we developed the data-reduction pipeline 
and fine-tuned the digitzer's performance and operation. This initial run consisted of 500 plates taken from a variety of different plate-series, all containing the 
open cluster Praeseppe (M44). We report that accurate photometry at the 0.1mag level is possible on the majority of plates, and demonstrate 
century-long light-curves of various types of variable stars in and around M44. 
\end {abstract}

\section{Introduction}
Astronomical source variability is poorly explored 
on ~1-100y timescales, where large scale systematic 
surveys are generally lacking. The Digital Access to a 
Sky Century @ Harvard (DASCH) project (\cite{grindlay2008}, in preparation; paper I) is designed to open the 
window on variability studies over a very wide range of 
timescales (~1d - ~100y) as recorded on the 
~500,000 Harvard plates. These were obtained full-sky over 
a century (1880 - 1985) from small telescopes in the 
southern and northern hemispheres and have been key 
for a number of fundamental astronomical discoveries. 
With our recent development of a very high-speed plate 
scanning machine \citep{simcoe2006}, it has become 
practical to digitize the full collection. In this paper, 
we report on the initial digitization and analysis 
software as developed from scans of a field centered 
on the Praesepe (M44) open cluster. The $\sim$600 plates 
scanned to include the cluster, and which cover 
a $\sim$20deg x 20deg field (with radially decreasing 
coverage), enable development of the astrometry 
and photometry analysis pipeline. We describe the key 
features of this analysis system and present initial results 
for astrometric and photometric accuracy and completeness 
that can be achieved with DASCH. Example light curves 
are shown for a variety of objects. A subsequent paper (\cite{tang2008}; paper III) extends the pipeline with additional processing 
steps (e.g. for source crowding and plate defect rejection) 
and reports on a generalized large-amplitude variable search 
for the M44 region.

\section{Digitization}
The DASCH digitizer \citep{simcoe2006} is capable of scanning one 11x17 inch or two 8x10 inch plates in less than a minute. The digitizer output is a set of 64 or 32 partially-overlapping frames which are then mosaiced  to form a single image of each plate. The design and workings of the digitizer are fully described by \citep{simcoe2006}, its essential features are
a precise (0.2 micron) X-Y stage carrying the plate; which is illuminated from below by a red LED array and imaged from above by a 4Kx4K CCD camera at 1:1 scale via a tele-centric 
lens assembly, the exposure is controlled by the LED pulse duration (7 ms average).
Digital images are captured via a 12 bit ADC,  which thanks to sensitive control of the exposure level is sufficient to record the full dynamic range of the photographic 
negative. The transmitted intensity scale is then inverted to give a positive image 
for further analysis. The camera pixels measure 11 microns which due to the telecentric (zero-magnification) lens are able to capture details finer than the emulsion grains. The design specifications for the digitzer are such that every salient detail of the original data is recorded, and therefore the job of scanning the plates will not need to be repeated. 

\section{Astrometric Solution} 
\begin{figure*}
\begin{center}
\includegraphics[angle=0,width=8cm]{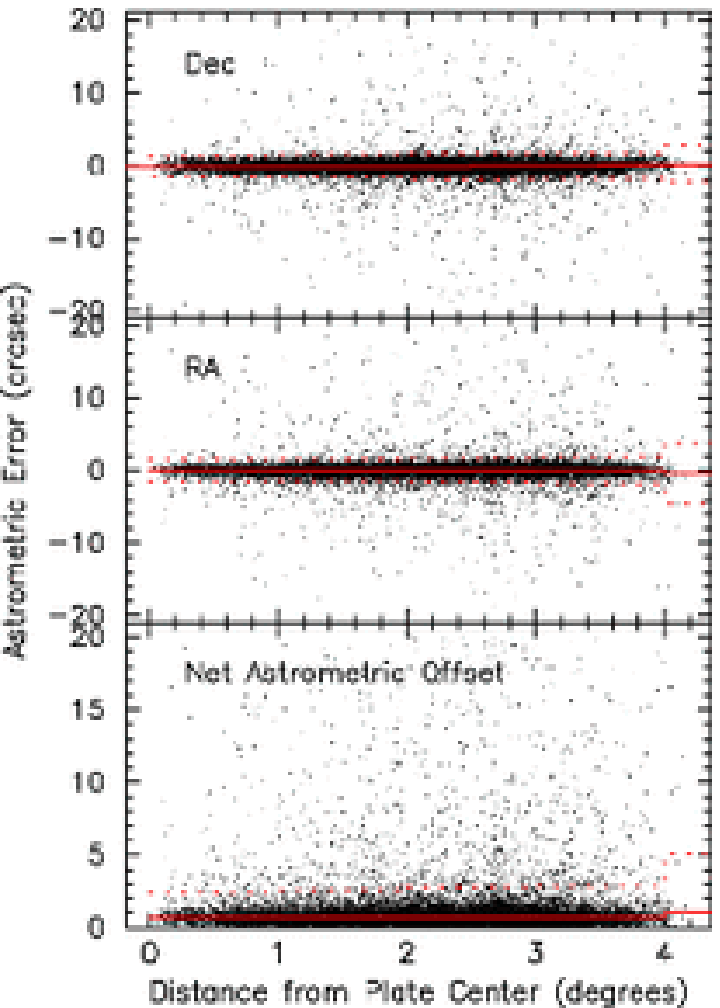}
\includegraphics[angle=0,width=8cm]{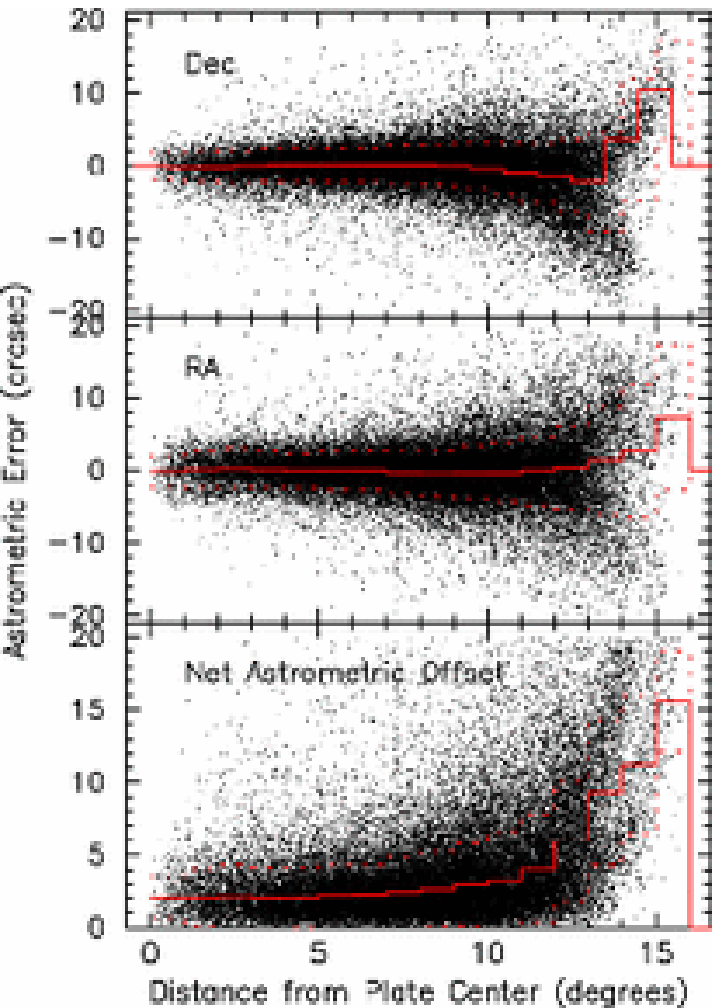}
\caption{{\bf Astrometric Plate Solutions.} Each plate is first fit with a tangent plane (TAN) projection to obtain accurate center and image-scale.
The TAN solution is used as a starting point for a TNX projection, which consists of TAN plus a polynomial distortion model to fit for optical aberrations.
This figure shows astrometric residuals following TNX fitting, for long (Left panel MC-plate) and short (Right panel: BM plate) focal length plate series. 
The X-axis is distance from plate center in degrees, and the Y-axes display the difference between the fitted and catalog positions, in arcseconds. Median and 
1$\sigma$ envelope are indicated by solid and dotted red histograms respectively. The points scattered at large offset from zero are generally mis-identified GSC stars.
}
\label{fig:wcs}
\end{center}
\end{figure*}

An automated procedure based upon routines in WCSTools (tdc-www.harvard.edu/software/wcstools/) first determines an approximate world coordintate system (WCS) for the plate. This determines the correct orientation of the plate, the image scale (''/pixel) and fits a tangent plane projection (TAN) to produce a rough ($10\sim20$ arcsec) celestial co-ordinate frame. The TAN solution is used a starting point for SExtractor (\citep{bertin}, hearefter $SE$) to detect all objects on the plate and assign initial RA, DEC coordinates. The SE catalog is matched against the GSC2.2 catalog with a 20'' matching radius, keeping only the closest match to each star. A rough photometric calibration is derived for the entire plate by fitting a quadratic to the GSC2.2 ($B$) vs instrumental magnitude, this calibration is used to reject incorrect matches (disagreement greater than 1 mag) to produce a clean calibration catalog.
Finally the calibration catalog is used as input to the IRAF task CCMAP, which fits a more accurate plate solution to the X,Y vs RA,DEC data. We use the TNX plate solution (http://iraf.noao.edu/projects/ccdmosaic/tnx.html) which consists of a tangent plane projection plus polynomial terms to model geometric distortions in the image. Distortions originate primarily in the original telescope/camera optics with only minor contributions from the DASCH scanner. Our lens introduces distortions  of order 1 pixel at the corner of the CCD frame. In the present work this is mitigated by overlapping the frames and averaging to create the full-plate mosaic image.  Following extensive testing of astrometric plate solutions it was found that a 6th order polynomial distortion model in X and Y is required to produce a solution free of systematic residuals, while increasing the number of terms produces no further improvement.  Examples of actual astrometric residuals for different plate-series are shown in figure~\ref{fig:wcs}, plotted as a function of distance from the center of the plate. The residual scatter (RMS) of the final plate solutions is of order 1 pixel, which corresponds to between 1 and 5 arcseconds depending on the plate-scale.

\section{Photometric Calibration}
The fundamental objective of DASCH is a precision ($\lesssim$0.1mag) database cataloging the changes in brightness of every resolved star in the sky brighter than about 17th magnitude, over 100 years \citep{grindlay2008}. Such a goal requires careful control of many potential sources of inconsistency. Following the assignment of an accurate astrometric solution to each plate, the transmitted intensity scale must be calibrated to yield stellar magnitudes on a standard system. Photometric calibration entails three distinct challenges: (1) detecting all real stars recorded on the plate (without introducing spurious objects); (2) measuring the brightness of every star in a consistent and optimized manner; and (3) removing the effects of uneven illumination (vignetting), development and emulsion irregularities. In addition the calibration must be consistent between separate plates and between plate-series, despite large variations in plate-scale, exposure-time, development, emulsion characteristics, telescope optics, and atmospheric conditions at the time of observations. In this paper we describe how all of the above variables are accounted for in the DASCH photometry pipeline.

\subsection{The Photographic Calibration Curve.}
\label{sect:characteristic_curve}

The sensitivity of photographic materials is traditionally 
described by the {\it characteristic curve} of log(Exposure) vs log(Density) where exposure is codified through a reciprocal arrangement of  exposure time and incident light intensity, density refers to a measure of number of developed grains per unit area, and hence the degree of darkening of the emulsion. The characteristic curve is normally 'S' shaped with a linear (in log-log space) portion in the middle, over which the aforementioned reciprocal relation is valid. At higher exposure values the linear regime breaks down, with increased exposure leading to an ever diminishing increase in density, a situation referred to reciprocity failure. 

Given the heterogeneous nature of the plate collection, and paucity of detailed records on emulsions, developers, and procedures at the different observing stations, it is safest to assume that the characteristic curve is different for every single plate. Contemporary photometric calibration was performed in a variety of ways. These included: (1) secondary exposures of standard star-fields (2) Multiple exposures obtained by opening and closing the camera shutter for precisely timed intervals, with the telescope drive turned off  in-between. So as to produce calibration sequences from the brighter stars on the plate, to be compared against the fainter science-program stars.  (3) The use of glass wedges and wire diffraction gratings to produce calibration sequences. These devices were placed over the objective-lens  to produce multiple images of the brightest stars on a plate. The diameters of these images are in an exact ratio to the primary images providing a calibration, as  described by \citep{chapman1913}. 
(4) Calibration wedges (small sections of the plate exposed by a lamp, through filters of known density) are present on a few percent of the plates.
None of these ingenious approaches are suitable for furnishing a uniform photometric calibration of the entire plate-collection. 

The obvious approach then, is to bypass the characteristic curve altogether and calibrate instrumental magnitudes measured from the plates directly, against a catalog of known magnitudes, via a non-linear fitting function. Previously this method was used with plate-data only for small regions of the sky where accurate magnitudes had been laboriously determined. However with the advent of all-sky (or large area) star catalogs such as the Hubble Guide Star Catalog  (GSC), Sloan Digital Sky Survey (SDSS), All Sky Automated Survey (ASAS), 2MASS and others, reference stars are available everywhere, with a sufficient density to guarantee coverage of any particular plate. In fact direct calibration via on-plate reference stars is superior to all other methods one could imagine. Provided the reference stars are densely distributed and cover the full range of star brightnesses recorded on the plate, local variations in sensitivity, illumination and point spread function (PSF) can all be mathematically fitted and corrected for. The fact that the reference stars lie on the plate means their light has followed exactly the same optical path and development conditions as their neighbors, removing all of the systematics inherent in using separate standard stars. In particular the effects of atmospheric absorption and atmospheric differential refraction are implicitly included in the calibration curve, provided the calibration star magnitudes are known in the same photometric band as recorded on the plate; or can be transformed into that band, as we describe below.

Following extensive research, the most accurate and complete reference catalog available for the full sky is the GSC2.2. Although itself derived from photographic plates 
the underlying photometric system of the GSC2.2 is tied to a set of photoelectric magnitude calibrators (the Guide Star Photometric Catalog- GSPC) which were observed for 
every plate in the POSS-II survey \citep{reid1991} (we cannot  use the GSPC because it is very sparse). The GSC2.2 magnitudes are provided for all stars in a blue ($B$) pass-band and for nearly all stars a red ($R$) pass-band also.  The GSC2.2 covers a large magnitude range $B$=1-19 with quoted uncertainties of $\sim$0.2 mag \citep{russell1990} and the 
bright end (B$<$10) coming from the Hipparcos satellite catalog \citep{hog1997}. The GSC-$B$ magnitude system turns out to be very close to the natural system of the majority of Harvard plates, which are primarily unfiltered blue-sensitive plates. The two GSC pass-bands also enable us to determine the wavelength response of each Harvard plate by fitting a color dependent term to the calibration curve. 

\subsection{Object Detection}
The first step in our photometry pipeline is to run Source Extractor \citep{bertin} on the astrometrically calibrated digital plate-image. $SE$ has many features that make it ideal for our purposes: It is fast, a crucial factor when one considers the software pipeline should eventually keep pace with the scanning: $SE$ takes 2--6 minutes for each 750 MB, 380 megapixel image on a single 3GHz CPU depending on the number of stars. Instrumental magnitudes are computed in fixed aperture, isophotal and Kron flexible aperture forms. Source radii at specified flux levels are written, providing an analog of the methods used by photometrists during the photographic era. Sky-background is modeled and available for inspection: a useful feature for examining the degree of vignetting and large-scale emulsion defects.  De-blending is performed to account for contamination of magnitudes by neighboring stars. We run SE with a low detection threshold (1.5$\sigma$) in order to be sure of detecting all potentially real objects. Due to the grainy nature of photographic emulsion and strong spatial gradients, standard image-noise models are not appropriate for detection thresholding, so we select the real objects after the fact, using a robustly derived limiting magnitude cut (described below).

\subsection{Fitting the Calibration Curve.}
The methods in use for CCD photometry all rely on the summation of pixel values belonging to each star. Under the assumption of a linear detector the derived flux (F) is directly proportional to the actual photon-flux received from the star. Calibration of CCD measurements is therefore is simply a matter of determining the constant of proportionality, which takes the form of a zeropoint-offset (Z) when the expression is formulated in magnitude units, as eqn~\ref{eqn:ccdmag}. 

\begin{equation}
M_{CCD} = Z - 2.5 Log_{10}(F)
\label{eqn:ccdmag}
\end{equation}

As described in section~\ref{sect:characteristic_curve}, photographic magnitudes follow a non-linear calibration curve, the shape of which depends  (in addition to the physical properties discussed in section~\ref{sect:characteristic_curve}) upon the measurement technique. For DASCH commissioning purposes we selected 3 methods: fixed aperture, isophotal and image-size. A more sophisticated profile-fitting technique will likely be implemented in future. However dealing with the spatially variable and peculiarly shaped PSFs found on the plates means that it will be limited to a subset of the data, while our present aim is to utilize as much of the collection as possible.

In  fixed aperture photometry, F is the sum of all pixel values lying in a circular aperture centered on the star's center-of-light (centroid). The functional form of this particular calibration curve has been determined by \cite{bacher2005}, in terms of 4 parameters which are in principle measurable from the image, or may be fitted for. The Bacher function enables very precise calibrations using only a limited number of reference stars, which need not extend much fainter than the magnitude at which there cease to be saturated pixels in the cores of the stars. Unfortunately fixed apertures mean that whatever aperture-size is chosen will only be the optimum choice for a single point along the curve. Apertures which are too large are dominated by sky-noise and stellar crowding, while at the bright end too small an aperture will capture only saturated pixels, leading to a flat (and therefore useless) calibration curve.

The most widely used method in the photographic era was measurement of the image diameter through a microscope, fitted with a reticle eyepiece or iris diaphram. The diameters of standard stars would be measured and compared to those of the program stars. This was (and still is, for example \cite{berdnikov2007}) performed by eye. Interpolating between closely spaced standard magnitudes, a level of accuracy approaching 0.1 mag is achievable by the best practitioners, although the work is painfully slow and requires a high degree of skill and experience. For comparison purposes we created an analog of this method, by calibrating our reference stars against their image diameters (in fact image area, to account for the non-roundness and spatial variation of the PSF), measured at a specified threshold. A previous example of microscope-measured image-diameters calibrated by on-plate standard stars via a fitting-function is provided by \cite{tinney1993} who used splines to parameterize the curve.  Our method is similar but of course the image-diameters are measured by software ($SE$) rather than micrometer.
  
Isophotal photometry (ISO) can be thought of as a combination of the preceding methods: a threshold is specified in terms of the local sky background, and all contiguous 
pixels above this threshold are assigned to the star. If any contiguous group of pixels has more than one peak, it is de-blended and the pixels assigned to the stars 
making up the blend. The instrumental magnitude in then derived from the summation of pixel values belonging to the star. This technique is implemented in SE and was inherited from the previous generation of digitizers such as the SuperCosmos Sky Survey \citep{hambly2001b}. A complete discussion of DASCH crowded field photometry is given by \cite{tang2008}.

\begin{figure}
\begin{center}
\includegraphics[angle=0,width=12cm]{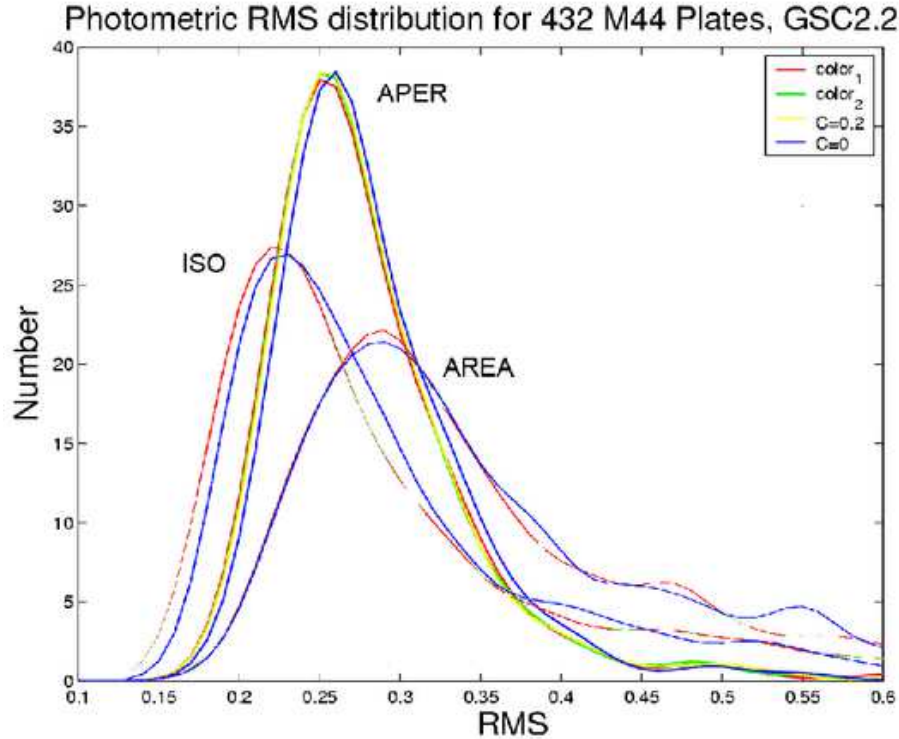}
\caption{{\bf Comparison of Photometric Consistency for ISO, APER and AREA methods.} Curves show the RMS-per-lightcurve in histogram form, for three photometric measurement schemes. The smallest RMS and hence the most consistent method is isophotal magnitudes fitted with the non-parametric {\it rlowesss} algorithm. Interestingly the peak of the distribution coincides with the stated uncertainty of GSC2.2  which was used as the reference catalog.}
\label{fig:compare_iso_aper_area}
\end{center}
\end{figure}

For our sample plates, we show the RMS distributions resulting from each of the three techniques in figure~\ref{fig:compare_iso_aper_area}. In order to calibrate the instrumental magnitudes
we need a fitting algorithm: in the case of fixed aperture there is a known functional form  \citep{bacher2005}, but for ISO and image-size there appears to be no universal function. The literature
contains many approaches, ranging from polynomials and splines, e.g. \cite{tinney1993} . A very sophisticated algorithm is described by \citep{russell1990} for the GSC, derived from photographic plates of the Second Palomar Schmidt Survey (POSS-II) that attempts to linearize the response by determining the characteristic curve from the star images themselves. 

\begin{figure*}
\begin{center}
\includegraphics[angle=0,width=14cm]{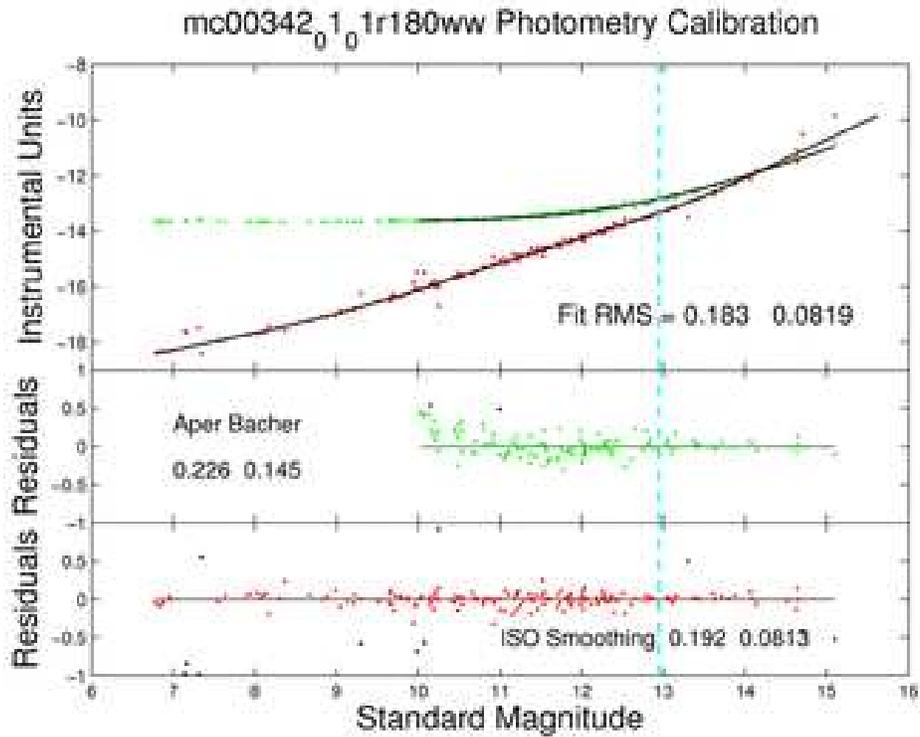}
\caption{Photometric calibration curves using WebDA photoelectric B magnitudes as reference stars. The upper panel shows fixed aperture magnitudes (red points), fit with 
Bacher's function \citep{bacher2005}, and isophotal magnitudes (green points) fit with the non-parametric {\it rlowesss} algorithm. Residuals from the two fits are plotted in the middle (aper) and bottom (iso) panels, in each case two RMS values are displayed: before and after removal of 3 $\sigma$ outliers. Note how the aperture magnitudes level-off
toward the bright end as the aperture becomes filled with saturated pixels, while isophotal magnitudes are well behaved across the full range of brightness.}
\label{fig:calibrationcurve}
\end{center}
\end{figure*}

After extensive testing with simple fitting forms, we found that no polynomial model works consistently well. We therefore turned to interpolation schemes which require no assumed function yet can be very robust against outliers. After some experimentation with box-car average and kernel density estimator approaches which allow the code to interpolate smoothly across gaps in the calibration sequence, we identified the statistical-analysis function {\it rlowesss} \citep{cleveland1981}, variants of which are incorporated in several widely used packages including ``R" and Matlab. The algorithm takes a series of X,Y points such as the calibration data ($B_{GSC}$ vs instrumental DASCH magnitude) and derives a smooth curve through the points. The algorithm is robust against outliers, applying an iterative sigma-rejection routine to identify and eliminate bad points from the fit . The (tabulated) value of the {\it rlowesss} curve is then used to convert (by interpolation from a very finely spaced grid) the instrumental ISO magnitude into a calibrated magnitude in the natural system of the plate. An example of a typical {\it rlowesss} fit to ISO magnitudes is given in figure~\ref{fig:calibrationcurve}, illustrating several important features: The quality of fit is seen in the residuals (lower panel), where the residuals are used to define an error envelope in order to attach uncertainties to each measurement. Extrapolation of the fit at the bright end is sometimes required if the plate contains too few bright stars to define a reliable curve. The limiting magnitude is determined by locating the point at which the scatter ($\sigma$) of the fit residuals begins to dominate over the slope of the curve, in practice we chose the point that is 2 $\sigma$ (two times mean std. deviation of the residuals) brighter than the faintest instrumental magnitudes returned by SE for real stars. Due to the grain inherent in photographic materials, which is often visible at size-scales similar to faint stars, the noise models used internally in SE (and indeed all other CCD photometry routines) are not strictly applicable and we therefore run the star-detection stage with a very low threshold to be sure of picking up all potentially real objects. False detections are dropped at the calibration stage, as most are not matched with reference stars and hence get ignored. Remaining false detections are outliers to the calibration curve and are removed by the rlowesss algorithm. All uncertainties are robustly derived by comparison of the final results against the reference magnitudes, by which we mean the fit residuals define the error distribution.

\subsection{Annular Photometry - Correcting for radially-symmetric vignetting and PSF degradation.}
\label{sect:annular}
The calibration curve is derived as a function of radial distance from the optical axis (physical center) of the plate. This enables allowance for the effects of vignetting and radially dependent PSF variations. The different plate series exhibit these effects to varying extents, which are greatest on those with a large plate scale. Vignetting cannot be removed by flat-fielding as with CCD images because the plate collection does not contain contemporary flat-field exposures, and the non-linear response precludes using flats constructed from the images themselves. The design of the telescopes and cameras provided a wide area of full illumination on the plates so that  vignetting is only noticeable toward the plate corners and is radially symmetric. 

PSF distortion is generally minor over most of the plate and tends to be mostly radially (but not azimuthally) symmetric.  Its effect on photometry is to change the fraction of light from a star that falls within a fixed aperture, such that fluxes tend to be under-estimated as one moves off-axis. Isophotal photometry alleviates the effect somewhat because all pixels above the threshold are included in the flux measurement, regardless of the shape of the profile. Using a low threshold ensures that flux-loss
occurs far out in the wings of the profile where it is less significant.  Both these effects change the shape of the calibration curve, and  also the limiting magnitude, since reduced effective exposure leads to diminished sensitivity. 

\begin{figure*}
\begin{center}
\includegraphics[angle=-90,width=14cm]{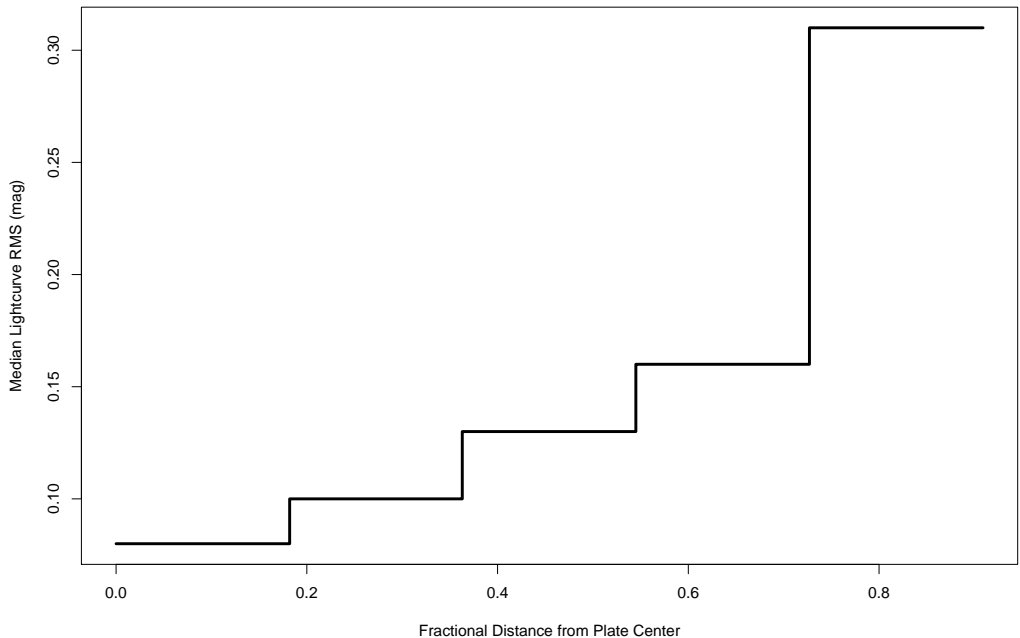}
\caption{{\bf Light-curve RMS vs Distance from Plate Centers.} Using the cluster M44 as a calibration region, light-curves were extracted for several hundred WebDA M44 cluster-members. Subsets of each lightcurve were selected according to distance from the center of the plate on which a star appeared. For example a given star might lie within 1$\degr$ of the center of 20 out of 300 plates. The median RMS for each radially-selected subset is plotted The plates displayed here are all  I-series which measure 9$\degr$x11$\degr$.}
\label{fig:radialscatterwebda}
\end{center}
\end{figure*}

The combined radial effects on photometry can be seen in figure~\ref{fig:radialscatterwebda} which plots the RMS distributions of light-curves constructed from measurements  located in specific radial bins. For this analysis we used M44 as a probe. The plates are all centered at different positions on the sky, so the cluster moves around in plate coordinates -enabling one to select points from each star's lightcurve containing only measurements made when it lay within a specific distance of the plate-center. Using only the M44 cluster-members meant we could calibrate our DASCH magnitudes against the WebDA  catalog (www.univie.ac.at/webda) which provides photoelectric  magnitudes for 300 stars with errors of  $\sim$0.02 mag. This enables very tight calibration, which is important in order to isolate the radial effects. The example data were drawn from 50 different plates, each one calibrated with a single calibration-curve to the WebDA stars. Due to the small size of M44 the cluster does not span a sufficient range in radial-distance for gradients within the sample to be significant. Figure~\ref{fig:radialscatterwebda} clearly demonstrates the effects of spatial variations as the light-curve RMS increases systematically with distance from plate-center.  The effect is seen in all plate-series with similar variation in radial plate offsets.   

Using a sufficiently dense reference star catalog (e.g. GSC se section 4.1) covering the entire plate, we construct and then fit the calibration curve in a series of concentric annular regions centered on the optical axis. Within each region the illumination and PSF are assumed constant. In addition a rectangular border around the plate, defined at 10\% of the minor axis is used to flag stars likely to be affected by edge-damage and uneven developer action (within a few mm of the plate edge, developer can penetrate horizontally into the emulsion, resulting in increased development). Following experimentation separately with the number and placement of the annular calibration regions we found that eight concentric equal-area bins are optimum in reducing the RMS. A ninth bin is the outer border (10\% minor axis) of the plate.

\subsection{Local Calibration - Correcting for localized spatial variations in plate sensitivity.}
\label{sect:localcal}
In addition to vignetting (dealt with in section~\ref{sect:annular}), there are sometimes irregularities in the emulsion on scales of $\sim$ 2 cm. Such features are evident to the eye and can be more clearly seen in the background maps and source-subtracted images output by SE. A possible origin is insufficient agitation during development and fixing of the plate. Agitation involves rocking the plate or the chemical-tank either by hand or  mechanically, in order to continuously redistribute the chemicals and prevent exhausted developer from remaining in contact with the emulsion. Any spatial differences in development result in changes to the characteristic curve, and hence the sensitivity and contrast, while incomplete fixing leads to foggy areas. On wide-field plates, local variations in calibration can also be arise from partial cloud cover during the exposures. 

By generating a map of the calibration residuals for each plate, we apply a third level of correction to the calibration process to remove irregular spatial variations. A regular grid of 50x50 positions in plate coordinates is defined covering the plate. At each position the average calibration residual is computed for all stars within a 0.5$\deg$ radius. The exact statistic is a clipped median (3 iterations, removal of 3 sigma outliers) including only stars brighter than the local limiting magnitude. The chosen reference catalog (GSC2.2) and sample radius provide 
$\sim$100 stars per grid-cell in the vicinity of M44, enabling residual maps to be generated on the required scale. The grid spacing is chosen to over-sample the spatial variations. the Once the map is generated, each star's magnitude is 
corrected by the clipped-median-residual of the nearest grid-point. From the above description it should be clear that the maps are inherently smooth due to the 
cell-radius being larger than the grid spacing, which ranges from 0.1$\deg$ on MC-plates to 0.5$\deg$ on RH plates. The physical spacing of the cell centers is 0.2 inches on the 8x10 inch plates. 

\begin{figure*}
\begin{center}
\includegraphics[angle=-90,width=8cm]{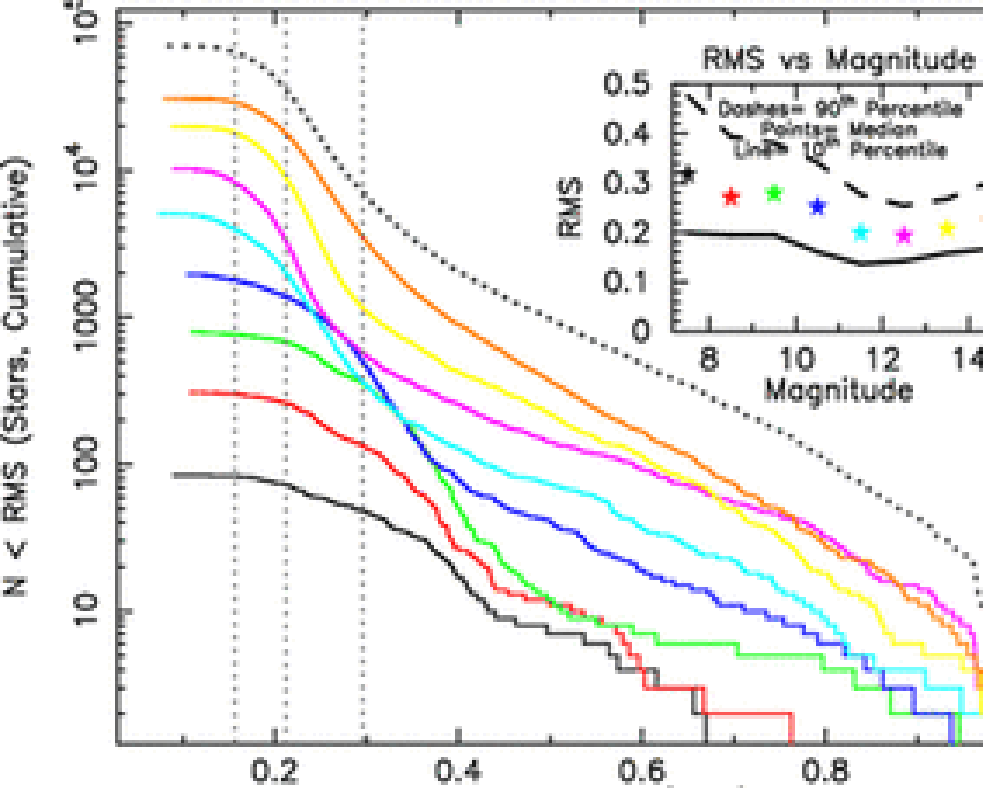}
\includegraphics[angle=-90,width=8cm]{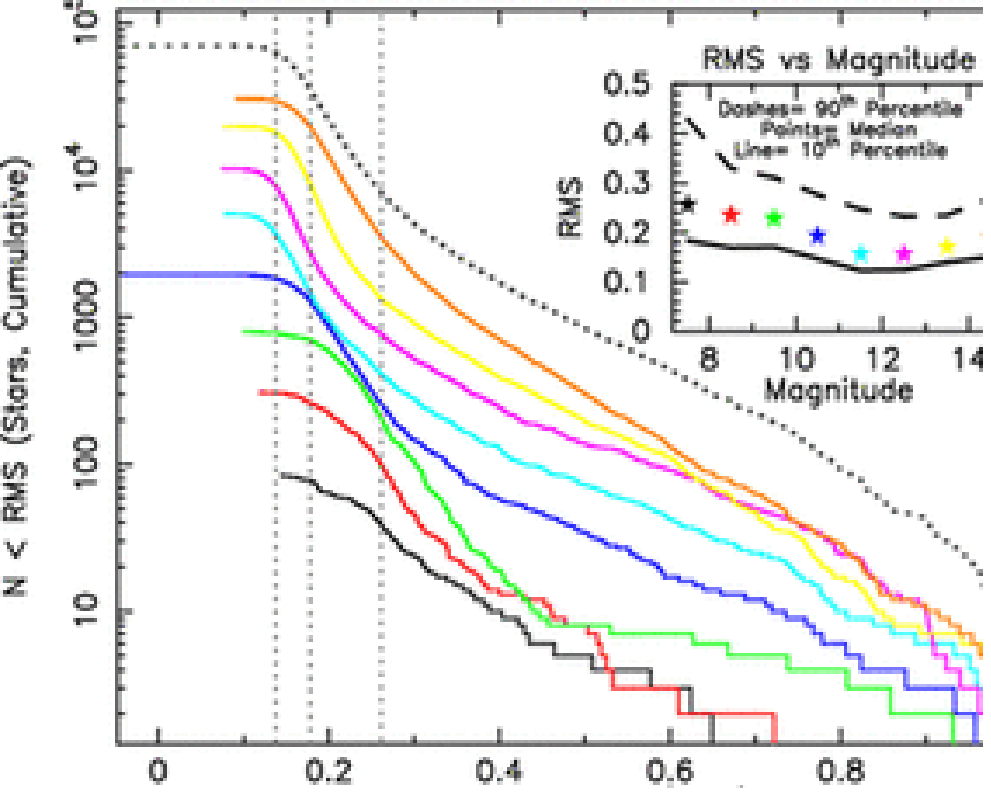}
\caption{Histograms showing light-curve RMS for 75,000 stars lying within 10$\degr$ of M44. The left panel displays isophotal photometry calibrated
using 9 annular regions per plate to correct for vignetting and radially-symetric PSF distortion. The right panel displays the same photometry after correction
of irregular spatial variations as described in section~\ref{sect:localcal}.  The RMS value is computed for each lightcurve after excluding unreliable points 
according to rules described in the text. The analysis is divided into eight magnitude ranges with the 10th, 50th \& 90th percentiles of RMS plotted in the inset panels versus magnitude using the same color-scheme as the histograms. Dotted histograms show the combined results, with percentiles plotted as vertical dotted lines. Following the correction-map procedure, clear improvement is evident in all histograms, and comparing the inset panels shows an across the board reduction in median RMS of 25-50\%. }
\label{fig:compareannularlocalrms}
\end{center}
\end{figure*}

\begin{figure}
\begin{center}
\includegraphics[angle=-90,width=18cm]{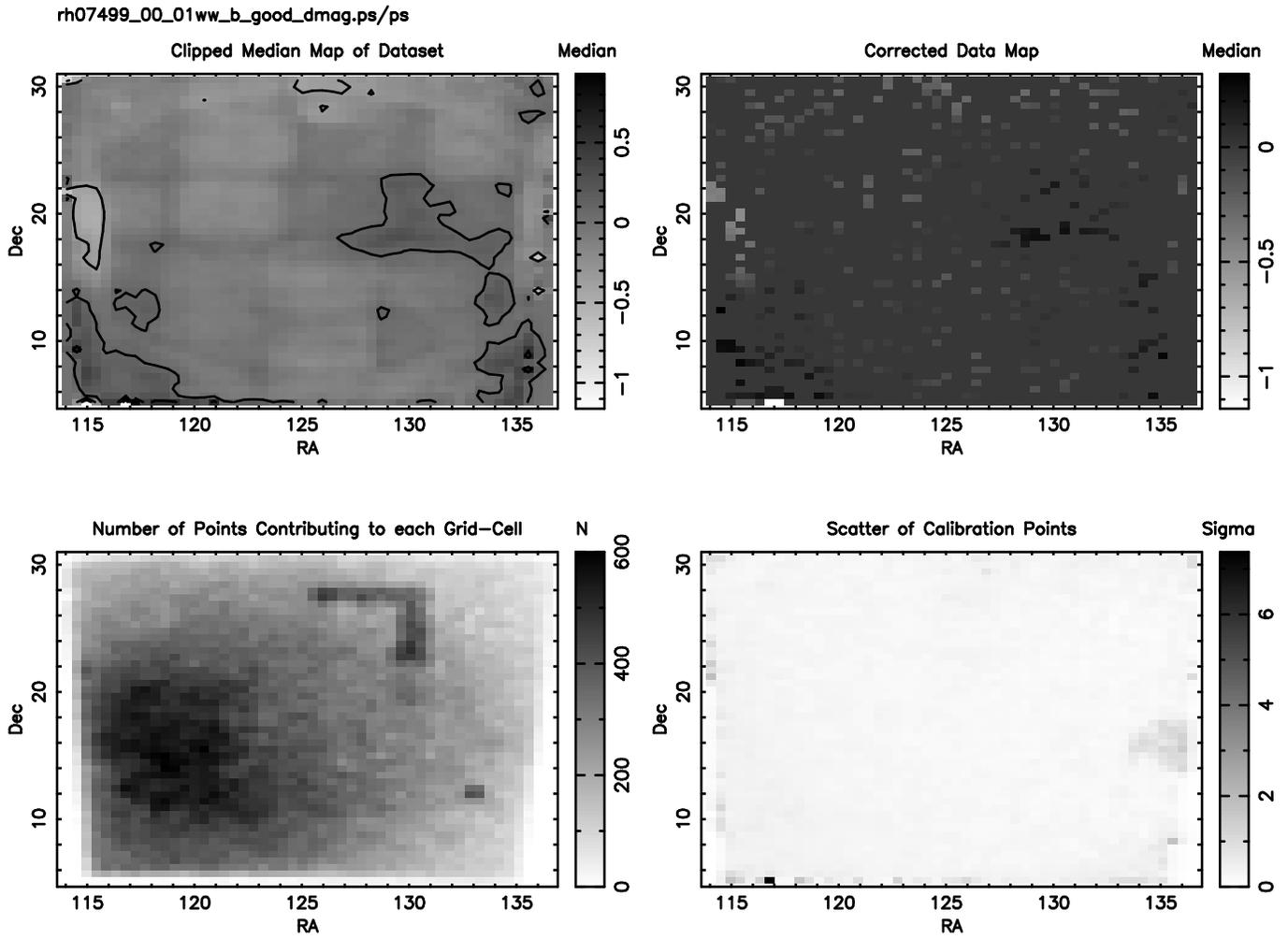}
\caption{{\bf Local photometric calibration map correction of irregular spatial variations.} The top left panel shows a map of calibration residuals (GSC2.2 - DASCH)
following annular-region fitting of the calibration curve, each pixel shows the clipped median of the residuals for stars within a 0.5$\degr$ radius. Bottom left panel shows
the number of stars included in the median calculation, while the bottom-right shows the RMS of the residuals. Finally the result of applying the calibration map is shown in the top-right panel. }
\label{fig:localmaps}
\end{center}
\end{figure}

Maps of the calibration residual, its inherent uncertainty, and number of points per cell are shown in figure~\ref{fig:localmaps}, which also includes a map of the final residuals, following correction. It can be seen that the annular calibration described in section~\ref{sect:annular} has removed all trace of vignetting, and the local calibration then drives the remaining spatial variations down by about 50\%. An analysis of light-curves generated from this photometry (see figure~\ref{fig:compareannularlocalrms}) shows the same level of improvement, demonstrating that spatial variations are handled effectively by the DASCH pipeline.

\subsection{Flags}
\label{sect:flags}
Certain plate defects can be difficult to completely correct for, but fortunately only a small fraction of stars are affected. Examples include emulsion irregularities, dirt and optical distortions. Fortunately these effects tend to be concentrated in the outer edges of the plates, which are most susceptible to handling, and lay furthest from the optical axis (relevant to distortion) of the telescope. In our photometric pipeline we flag stars for the following factors:
distance from plate center, proximity to plate edge, quality of {\it rlowess}  calibration-curve fit, quality of local-map correction ($\sigma$ and number of local reference stars used), local limiting-magnitude, SE flags (blend, neighbor-contamination), existence of known near-neighbors. All of these conditions are evaluated on a plate-by-plate and star-by-star basis, enabling very specific filtering of the DASCH database to investigate systematic effects and to include/exclude data-points for scientific analysis.  Further advances in filtering will be described in Paper III.
In this paper most numerical results include only data-points meeting our default quality thresholds: 1) Star not within 10\% minor-axis distance of plate-edge. 2)Total RMS of calibration-curve residuals less than 1 mag. 3) Magnitude brighter than 2$\sigma$ limiting magnitude for relevant annular bin. 4) Plates without multiple exposures, gratings, prisms, filters.
Generally in plots all points are shown, where possible smaller plot-symbols indicate flagged points. 

\section{Color-Response of Photographic Emulsion.}
The majority of the Harvard plate collection are blue sensitive, unfiltered plates. A small fraction used filters to produce red and yellow sensitive measurements although details of the wavelength response of the various emulsions are not available. In order to generate consistent and precise magnitudes for temporal analysis it is essential that changes in color-response from one plate to another be well understood, otherwise small brightness variations could be masked (or mimicked) by calibration shifts. 

Any given plate will yield magnitudes on a unique ``natural system'' defined by the combined wavelength-dependent transmission properties of the atmosphere, telescope and camera optics, and the emulsion response. Fortunately all of this can be reduced to a single co-effficient: the color-term $C$, which appears in equation~\ref{eqn:color}
relating magnitudes on the standard system, $M$, to the plate's system $m$, where $M_1$ and $M_2$ are two standard pass-bands close to $m$. This simple form assumes that a given star's incident spectrum is approximately linear over the effective bandpass of the system, and can therefore be completely described by the photometric color index ($M_1$--$M_2$). Since real stars have spectra of differing shapes, the chosen standard passbands should ideally bracket $m$ in effective wavelength. This discussion is equally applicable to CCD photometry and explains why the best results are obtained when standard stars are chosen to be of similar color to program stars.

\begin{equation}
m = M_{1} + C (M_{1}-M_{2}) 
\label{eqn:color}
\end{equation}
  
\begin{figure}
\begin{center}
\includegraphics[angle=0,width=12cm]{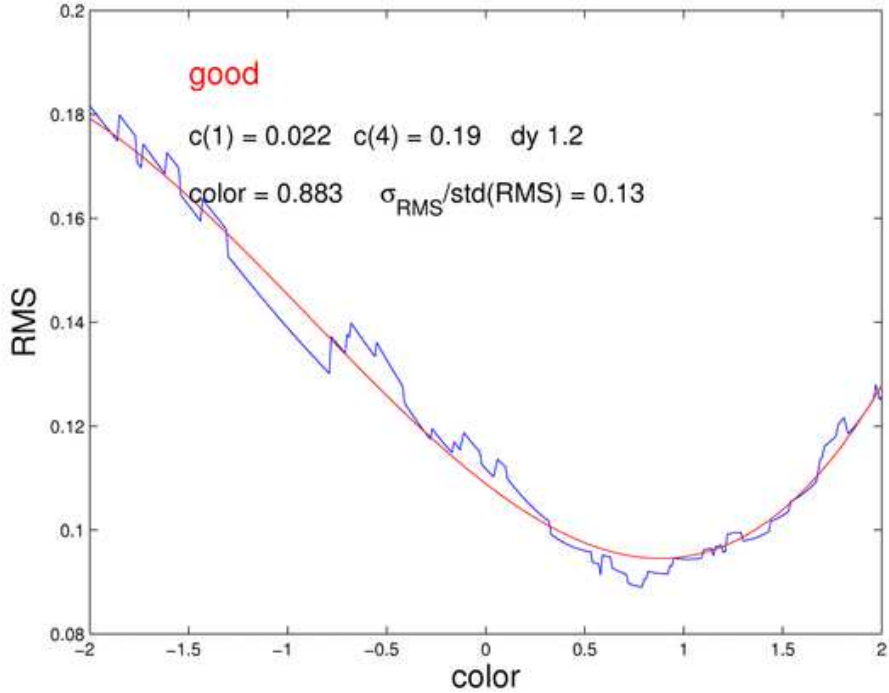}
\caption{{\bf Fitting the Color-Term.} The color-response of each plate is determined by stepping through a series of values of color-term $C$, repeatedly fitting the calibration curve: in this case $M_{ISO}$ vs $V+C(B-V)$, so as to locate the $C$ value producing the minimum RMS in the residuals. The reference stars are WebDA photoelectric magnitudes for M44 cluster members and the minimum scatter occurs for $C$=0.88, meaning that $M_{ISO} = V_{WebDA} + 0.88 (B-V)_{WebDA}$ hence this is confirmed to be a blue sensitive plate with an effective passband very close to Johnson-B.}
\label{fig:colortermfit}
\end{center}
\end{figure}

For every digitized plate then, the first task is to measure how its natural magnitude system compares to the standard one, by determining $C$. Our procedure was to step through a series of closely spaced $C$ values. For each trial value we converted the reference stars' magnitudes according to equation~\ref{eqn:color} and then fitted the instrumental plate-magnitudes against $M'$. For example when using reference stars from  the GSC2 catalog, $M_{1}$ and  $M_{2}$ are Johnson $B$ and $R$ magnitudes respectively, and we have Equation~\ref{eqn:color2}.  When the correct color term is applied, the scatter of residuals (Std deviation of $\Delta m$ computed for every star)  about the calibration curve reaches a minimum, as demonstrated in figure~\ref{fig:colortermfit}. 

\begin{equation}
\Delta m = m_{DASCH} - B = C (B-R) 
\label{eqn:color2}
\end{equation}

\begin{figure}
\begin{center}
\includegraphics[angle=0,width=12cm]{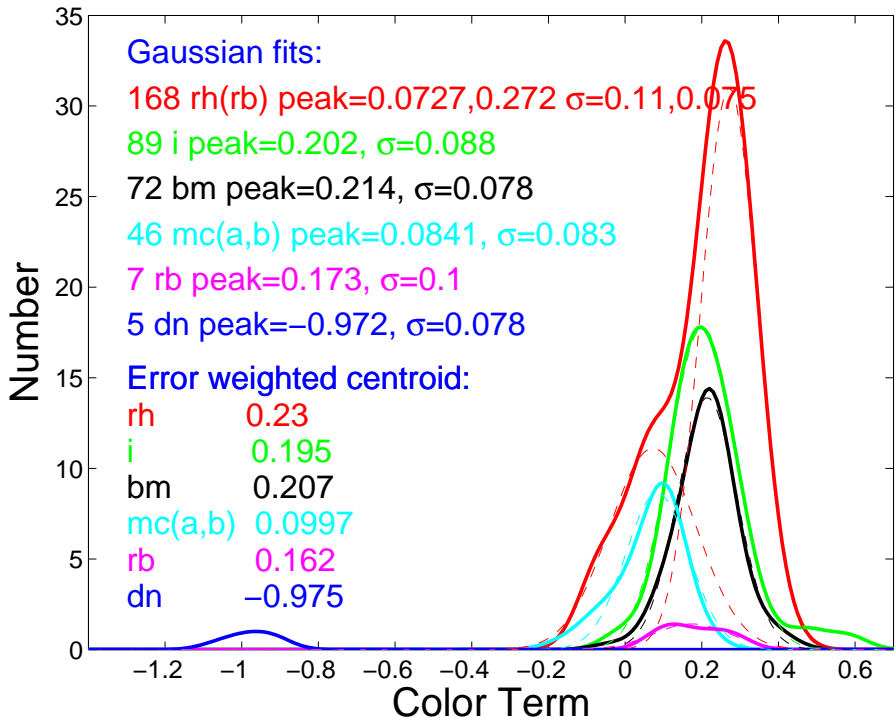}
\caption{{\bf Color-Term $C$ by Plate Series} The color-terms were obtained by minimizing the scatter in the calibration curve for each plate, 
relative to Johnson B, B-R magnitudes. Therefore in the above histogram plot,  B magnitude corresponds to $C$=0, which is very close to the result obtained for most of our sample.
Each plate-series defines a distinct distribution which is well described by a single gaussian, parameters from these fits are given in the legend. Five DN plates observed through a red filter were included for comparison, these appear at $C\sim$-1, which corresponds closely to Johnson R, confirming the reliability of our technique.}
\label{fig:colorbyseries}
\end{center}
\end{figure}

The color term for each plate is determined using a region near the center where the calibration curve is most cleanly defined.
Color response may vary slightly across a plate due to differential airmass, which would be most pronounced on wide-field plates. The effect is dealt with in Paper III \citep{tang2008}.
During this commissioning project we analyzed $\sim$500 mostly blue plates from 6 different plate series, spanning 100 years, purposely including a few yellow and red filtered plates. Reference stars were drawn from the GSC2.2 which provides B and R magnitudes accurate to $\sim$0.2mag. For a more precise comparison analysis over a small area centered on M44, we used the WebDA (www.univie.ac.at/webda) catalog which provides precise (0.01) magnitudes in B \& V for about 300 M44 cluster members. The results of the color analysis show systematic differences in $C$ between the different blue-sensitive plate series, at the level of about 0.1 mag. This translates to an order of magnitude smaller systematic shifts between the natural magnitude systems, since the majority of M44 cluster members have (B-V) colors less than 0.5 mag, resulting in typical $\Delta$m$\sim$0.05. These results are displayed in figures~\ref{fig:colorbyseries}  showing how the color-terms are distributed according to plate series. It was initially expected that early plates might be significantly bluer as panchromatic emulsions became widely available only in the 1900s, and that different batches might exhibit different color-sensitivities, not to mention the largely unknown issue of filters. It comes therefore as a surprise that histograms of the number of plates versus $C$ (figure~\ref{fig:colorbyseries}) show only small differences between plate series. The width of each histogram is of order 0.1 mag, which is comparable to the differences between their modes. The small shifts are real however and the RH histogram shows an intriguing `shoulder' which upon closer examination
is due to a change in color-response in 1932.

\section{Lightcurve Extraction}
Lightcurves are the ultimate purpose of DASCH and also provide an independent way to verify the quality of the astrometry, photometry and metadata.
The photometric calibration pipeline described above produces a catalog of stellar magnitudes and positions over time. To generate the light-curve of a given star, its brightness history as recorded in the set of plate-catalogs is ordered against observation time as recorded in the original logbooks. The information held in the logbooks consists of a one-line handwritten entry for each plate, noting the date and time of observation, exposure time, RA \& Dec of the plate center,  Hour Angle (HA), plus comments on any filters, prisms or gratings used, weather conditions and so on. From 1880 until 1953 the start and stop time of every exposure was recorded in local sidereal time, after which a mixture of local time (e.g. EST) and UT was used. Prior to 1925 dates were recorded as local calendar date at the beginning of the night. From 1925-1953 two local dates are given, corresponding to the start and end of the night.  In order to use the logbook records, they were digitally photographed and the images transcribed by a data entry company. Finally the meta-data files were converted to a common system; the Modified Julian Date (MJD) by software designed to determine the time system in use, and to which date the sidereal or UT time-stamp belonged. This is a fairly complex problem because the observations were made at several different latitudes, and finding the correct date requires calculation of the sidereal time of local midnight. In principle the values of Time, RA/Dec and HA should all agree,  so the software makes a comparison and if a conflict exists, `votes' the most probable solution and flags the entry for human investigation. For the purposes of temporal analysis it is more useful to compute the Heliocentric Julian Date (HJD), which corrects times to an imaginary clock located at the solar-system barycenter. This is required to correct for a shift in photon arrival time of up-to $\pm$8 minutes, which would be a significant source of error in the study of short  period phenomena such as compact eclipsing binaries.

Light-curves for every resolved GSC2.2 star brighter than B=17 in the region of sky covered by the M44 plates were extracted using a highly optimized piece of software based on the {\em Starbase} package (a command-line driven relational-database for astronomy: http://cfa-www.harvard.edu/~john/starbase/starbase.html).  The code relies on pre-sorting and indexing the photometry catalogs; each 
one is read only once by the search tool, which identifies all target stars that lie on the plate and takes advantage of the index to rapidly locate them (without having to examine every entry). If an expected target star is not found in a particular plate-catalog, then the local limiting magnitude for its position is returned, to be treated as an upper limit.
Execution time for 500 plates each containing an average of 42,000 stars, is $\sim$1 second per light-curve, for a single-CPU-thread.  Further optimizations are described in Paper III  which enable rapid processing of thousands of plates.

\section{Comparison between DASCH and GSC2.2 Magnitudes.}
Careful examination of the outliers in our photometric calibration-curves revealed three classes of astronomical object :  Variable stars, Galaxies and Blends.
Of these, variable stars obviously do not fall on the calibration curve unless they happen by chance to be at their catalog brightness on a particular plate. 
Galaxies do not conform because of their non-stellar profiles, the calibrations for aperture and image-size magnitudes do not apply to extended objects.
Isophotal and Kron (flexible aperture) magnitudes formally apply to any object, however changes in image-scale and photometric depth cause the 
galaxy magnitudes derived from different plates to disagree with one another. In deeper images more of the galaxy gets above the detection threshold, and 
hence the isophotal area expands and observed flux increases. Blended stars and those with near neighbors can yield unreliable magnitudes that may
mimic real variables as seeing, plate-scale and depth change between plates. Happily there are existing all-sky catalogs (e.g GSC2.2) that go deeper than most of the 
Harvard plates and are largely complete down to our typical limiting magnitudes. Prior knowledge of positions and magnitudes enable one to predict which stars 
will suffer from neighbor effects in a given plate.
   
\section{The DASCH Photometric Catalog}
We can take advantage of the hundreds of independent measurements of each star to average-out systematic and statistical errors. In this way for non-variable stars
DASCH has the potential to create an all-sky photometric catalog. For variable stars the median brightness will be more representative than that measured at a single instant, and the variability amplitude will be known. Among the applications of such a catalog are an improved GSC for observatory and spacecraft operations, as well as a list of standard stars for general calibration use. It has been recently discovered that a significant fraction (5\%) of the standards in current use are in fact long-term variables \citep{vogt2004}, DASCH will define a large set of constant stars proven to be good standards over our hundred year baseline. 

\begin{figure}
\begin{center}
\includegraphics[angle=0,width=12cm]{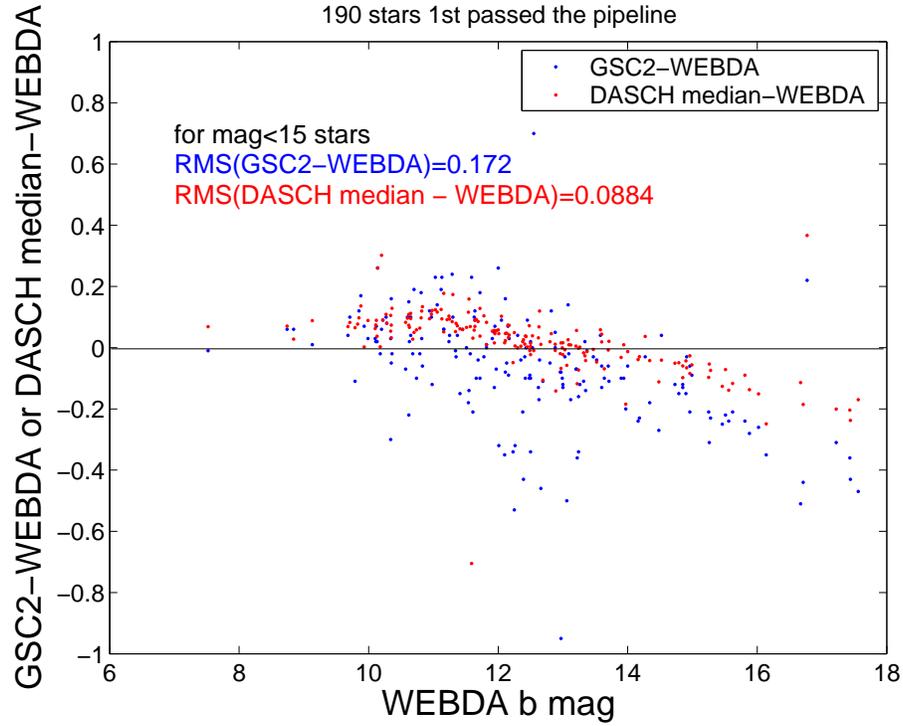}
\caption{{\bf Comparison of DASCH Median and GSC2.2 against WebDA Magnitudes.} The RMS of the discrepancies indicate the improvement in photometric precision made possible by averaging magnitudes over many plates. The raw values are given in the legend. After iterative outlier clipping (blended stars, false matches) the following values were 
obtained: RMS(DASCH - Webda) =  0.08 mag, a major improvement over RMS(GSC2 - Webda) =  0.22 mag, which itself closely matches the quoted accuracy of the GSC2.}
\label{fig:dasch_vs_gsc}
\end{center}
\end{figure}

For the DASCH commissioning project we computed the median magnitudes for 75,000
GSC2.2 stars within 10$\degr$ of M44 in order to refine our own calibrations. Median magnitudes were computed using only good measurements (flag specifications given in section~\ref{sect:flags}), for stars having at least 10 such points in their lightcurve. As a benchmark we compared the DASCH (median) and GSC2.2 magnitudes against WebDA photoelectric B magnitudes which have a precision of order 0.01mag. The overall RMS of $B_{WebDA}-B_{DASCH}$ is 0.08 compared to a much larger value of 
0.22 for  $B_{WebDA}-B_{GSC}$, in both cases after removing obvious outliers, this result is shown in figure~\ref{fig:dasch_vs_gsc}.
A large improvement in plate-calibration is possible by using the DASCH median-magnitudes in place of GSC2.2 to refine the individual curves, as demonstrated in 
figure~\ref{fig:dasch_vs_gsc}. Two improvements are evident: firstly the scatter of points about the {\it rlowess} fit is dramatically tightened; secondly the large-amplitude outliers are almost completely removed. It is clear therefore that iterating our calibration pipeline leads to impressive gains in both consistency and absolute precision. Further development along these lines is presented in Paper III.

\section{Demonstration Science Lightcurves}
\begin{figure*}
\begin{center}
\includegraphics[angle=-90,width=16cm]{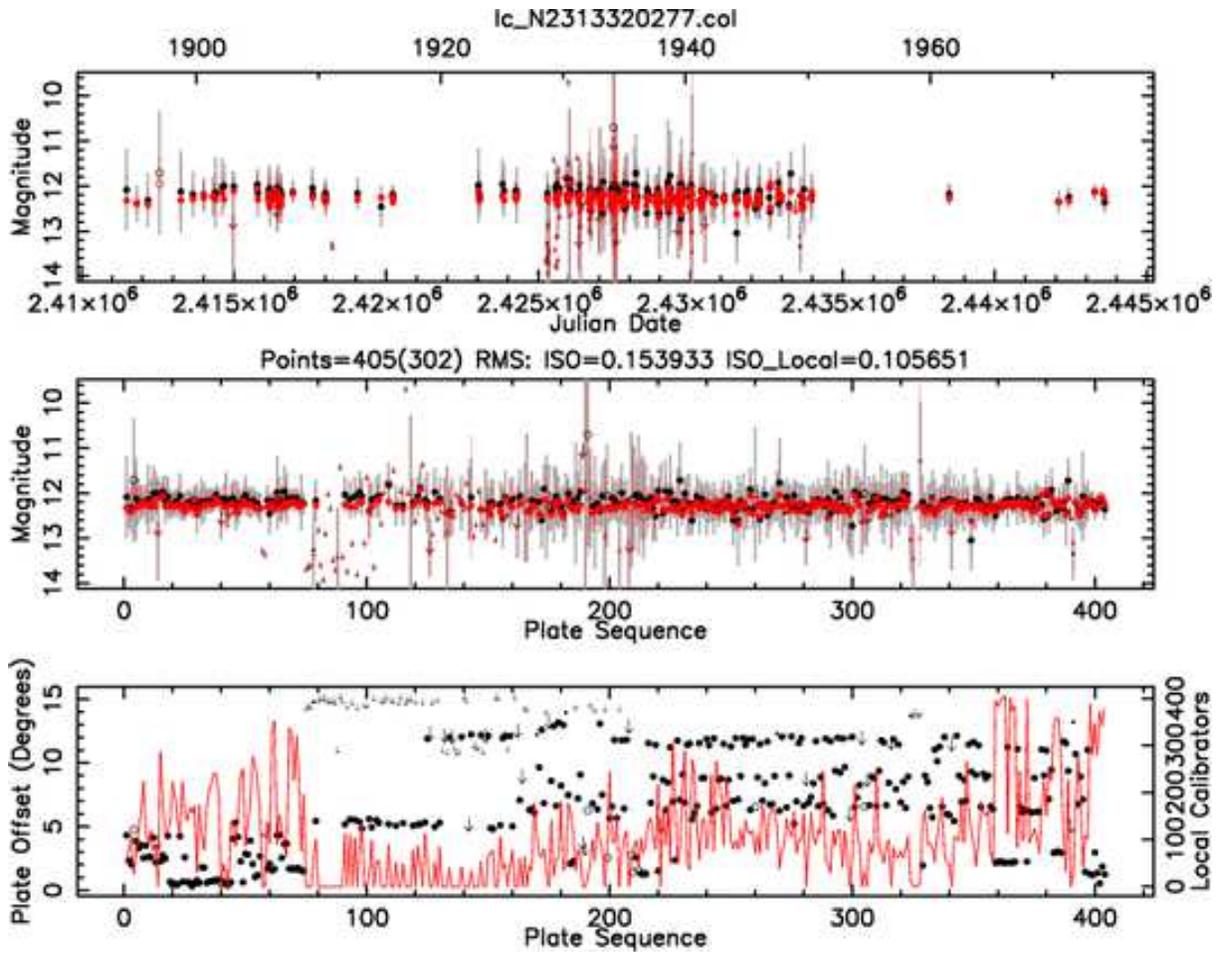}
\caption{Lightcurve of a constant star demonstrating about $\pm$0.1mag photometry over 400 plates, that span 100 years and 6 different plate-series.  The star's magnitude is plotted against time (upper panel) and chronological plate sequence (middle panel). Large solid circles denote good data points (of which there are 302), while open circles indicate a larger than usual scatter in the calibration for that point. Arrows denote upper limits derived for plates which did not go deep enough to detect the star. Smaller versions of these symbols denote observations where the star was very close to the plate-edge. Two versions of the lightcurve are shown: black points denote a calibration that accounts for radially-dependent effects such as vignetting, while the red points incorporate an additional step correcting for irregular spatial variations in plate sensitivity. The bottom panel shows for each measurement how far the star was from the plate-center, and the number of local calibrators used to correct for spatial variation. }
\label{fig:ConstantStar}
\end{center}
\end{figure*}

\begin{figure*}
\begin{center}
\includegraphics[angle=-90,width=16cm]{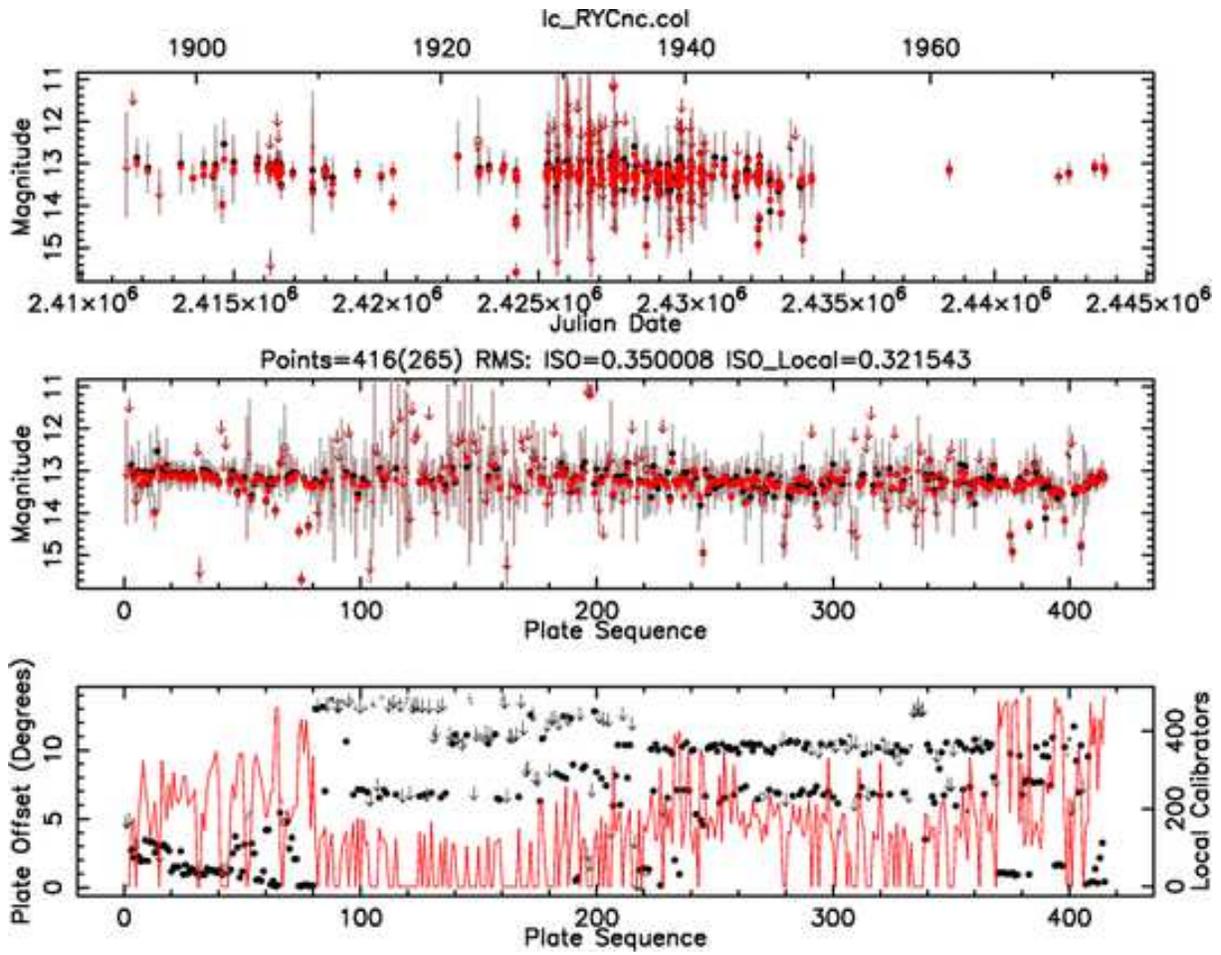}
\caption{Lightcurve of the eclipsing binary RY Cancri. This lightcurve contains 417 measurements of brightness obtained over 88 years. Note the occasional faint points below the rest, which correspond to eclipses of the primary star by its fainter companion.}
\label{fig:RYcnc}
\end{center}
\end{figure*}

\begin{figure}
\begin{center}
\includegraphics[angle=-90,width=12cm]{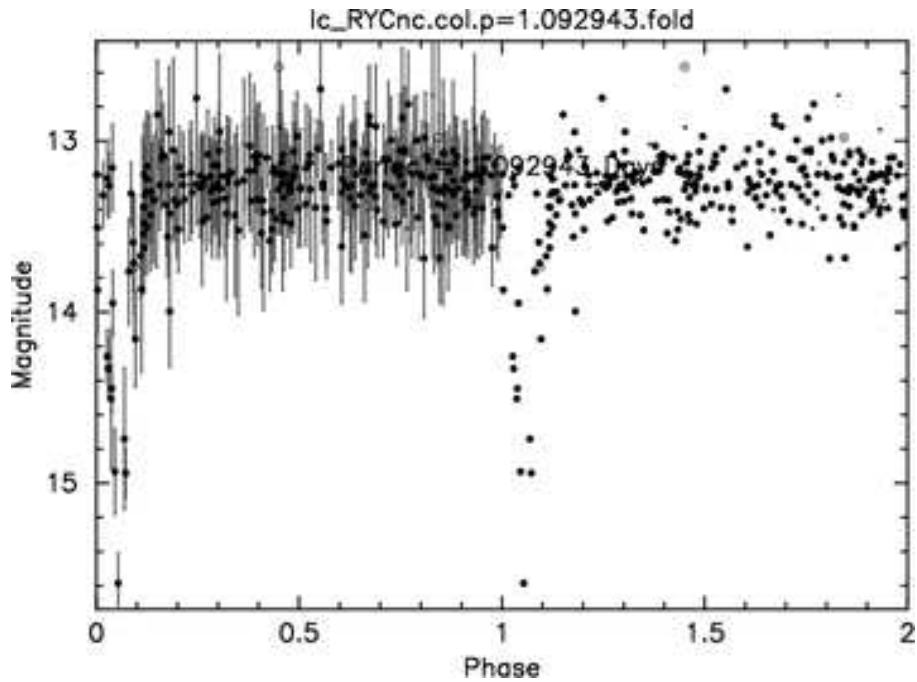}
\caption{RY Cnc  lightcurve folded at the known binary  period of 1.092943 days. }
\label{fig:RYcncFold}
\end{center}
\end{figure}

\begin{figure}
\begin{center}
\includegraphics[angle=-90,width=16cm]{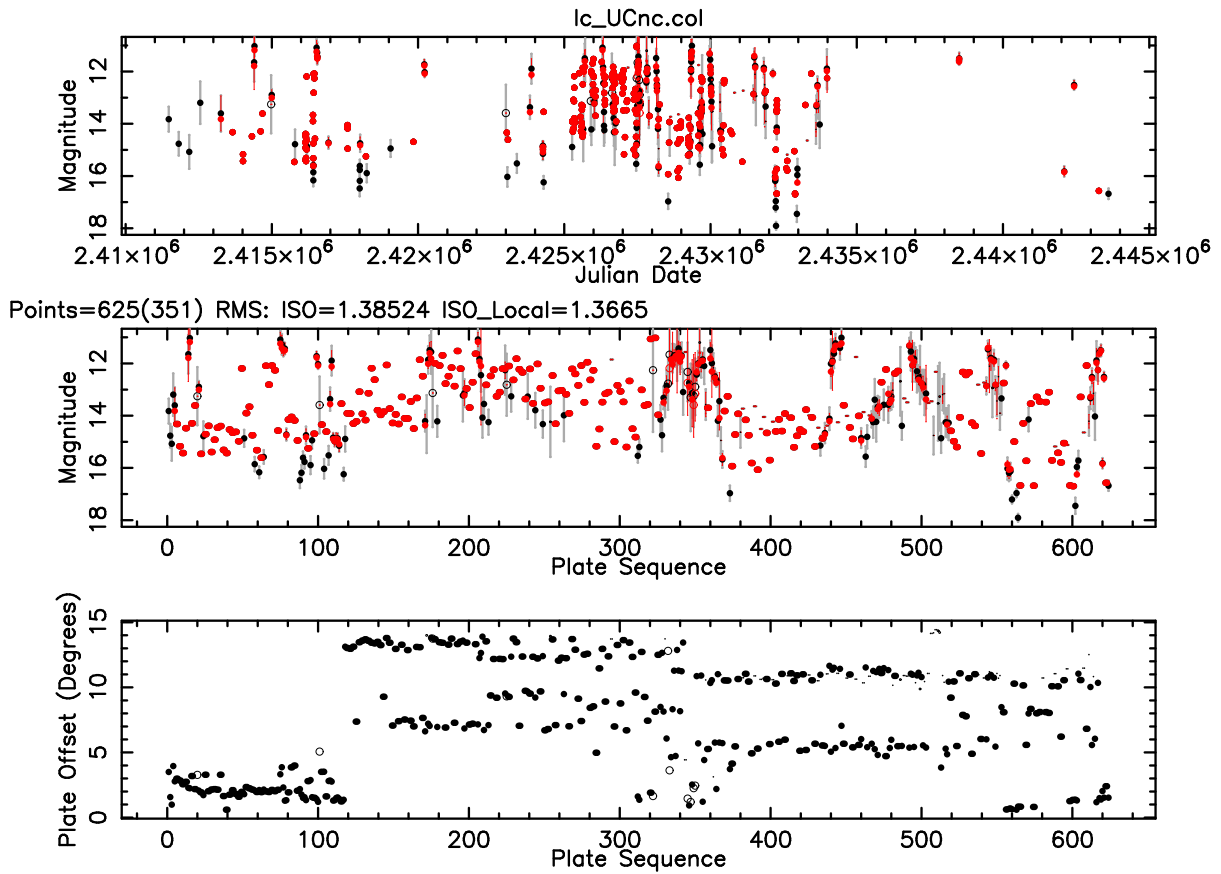}
\caption{U Cancri, a semi-regular variable of the Mira type. This is an example of a red-giant star that is slowly pulsating. The star brightens and fades by over 4 magnitudes  roughly every 300 days. During its minima U Cancri is not visible on most of our  plates because it gets too faint, as can be seen from the many upper limits (arrow symbols). }
\label{fig:UCnc}
\end{center}
\end{figure}

\begin{figure}
\begin{center}
\includegraphics[angle=-90,width=12cm]{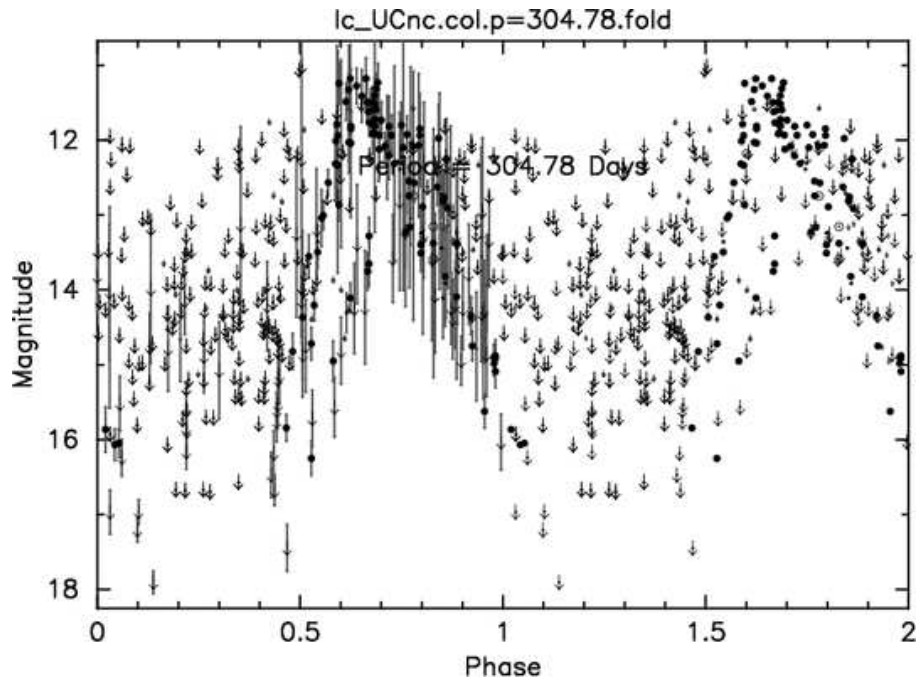}
\caption{U Cancri lightcurve folded at a period of 304.78 days, which was measured from the DASCH lightcurve shown in figure~\ref{fig:UCnc}. Because the star's pulsations are not perfectly regular, the points from different cycles don't line up exactly;  at least 2 distinct rising portions are visible, beginning at phase 0.5. }
\label{fig:UCncFold}
\end{center}
\end{figure}

\begin{figure}
\begin{center}
\includegraphics[angle=-90,width=16cm]{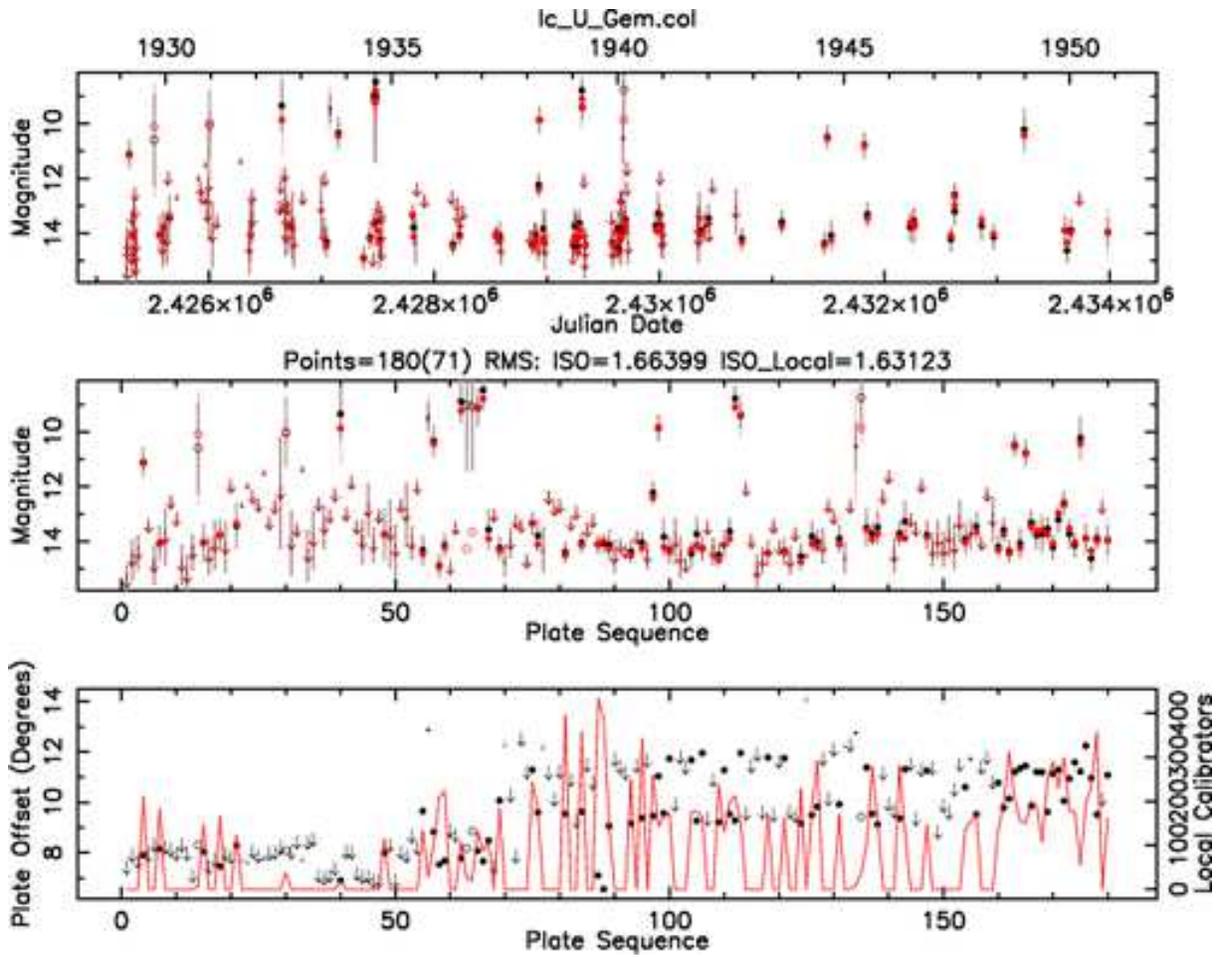}
\caption{{U Gem, the most famous recurrent dwarf nova.  The DASCH lightcurve shows a scattering of bright points corresponding to DN outbursts.  Upper limits appear for plates that did not reach deep enough to detect the system in its quiescent state. }}
\label{fig:UGem}
\end{center}
\end{figure}

\begin{figure}
\begin{center}
\includegraphics[angle=-90,width=16cm]{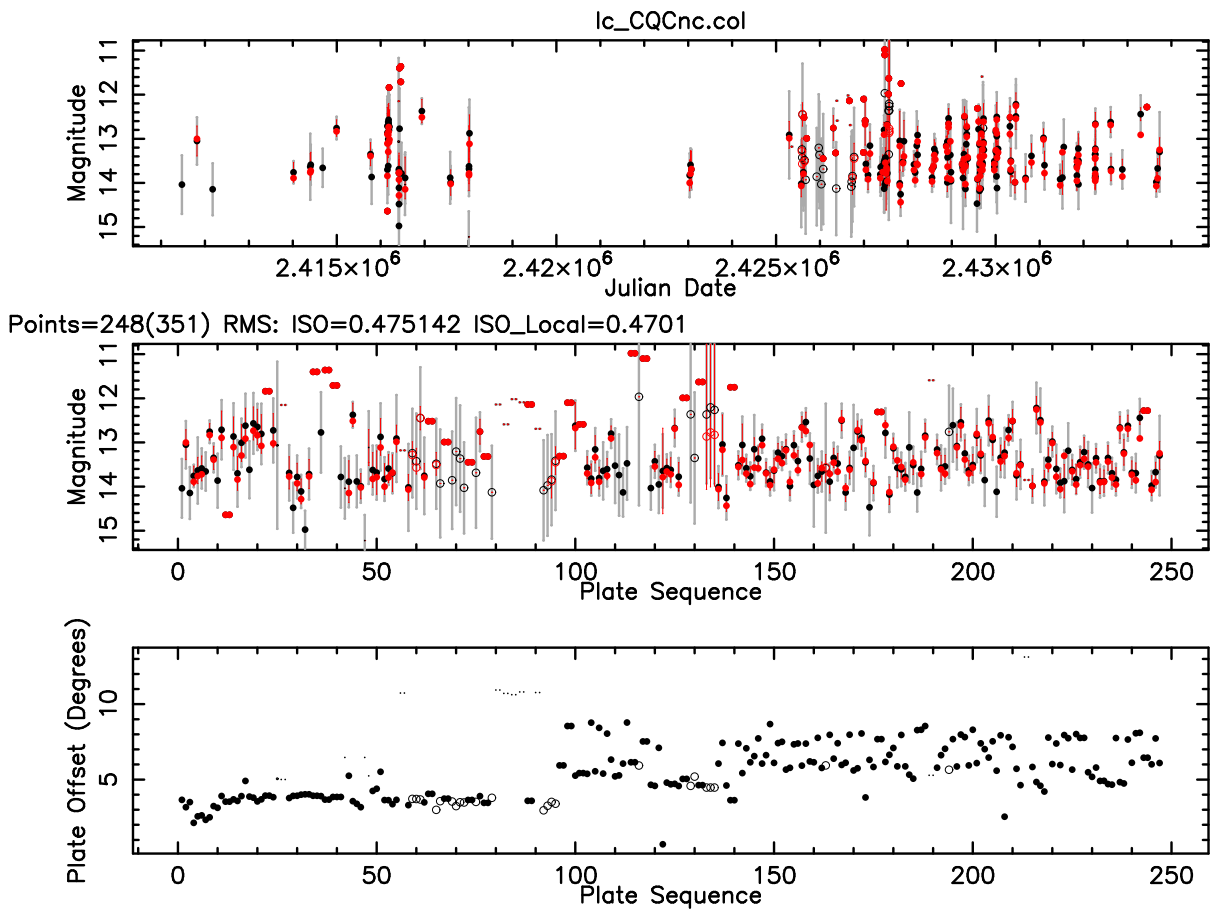}
\caption{CQ Cancri DASCH lightcurve}
\label{fig:CQCnc}
\end{center}
\end{figure}

\begin{figure}
\begin{center}
\includegraphics[angle=-90,width=12cm]{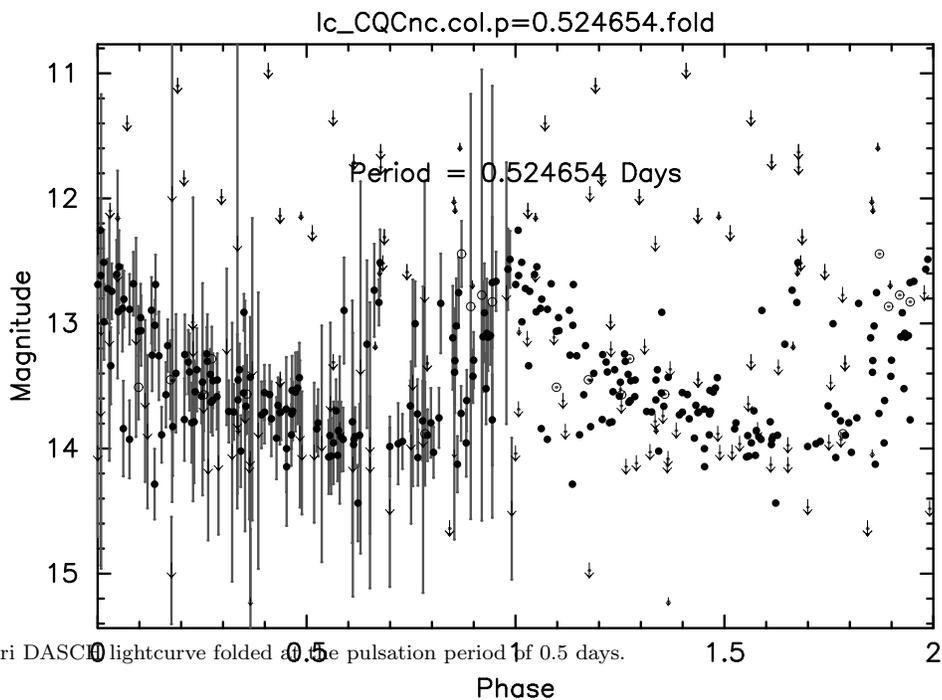}
\caption{CQ Cancri DASCH lightcurve folded at the pulsation period of 0.5 days.}
\label{fig:CQCncFold}
\end{center}
\end{figure}

Many variable stars are known within a few degrees of the well-known open cluster M44, including dwarf novae (DNe), eclipsing binaries, and various classes of pulsating stars. The examples in this section are intended to demonstrate the value of DASCH in a range of applications. In figures~\ref{fig:ConstantStar} -- \ref{fig:CQCncFold} the star's magnitude is plotted against time (upper panel) and chronological plate sequence (middle panel). Large solid circles denote good data points, while open circles indicate a larger than usual scatter in the calibration for that point. Arrows denote upper limits derived for plates which did not go deep enough to detect the star. Smaller versions of these symbols denote observations where the star was very close to the plate-edge. Two versions of each lightcurve are shown: black points denote a calibration that accounts for radially-dependent effects such as vignetting, while the red points incorporate an additional step correcting for irregular spatial variations in plate sensitivity. The bottom panel shows for each measurement how far the star was from the plate-center (points) and how many local calibrators were used (red line).

The first example is a star that does not appear to vary. Such constant-brightness stars enable sensitive determination of various systematic effects and provide a completely independent measure of uncertainties. In figure~\ref{fig:ConstantStar} is the lightcurve of such a star demonstrating about +/-0.1mag photometry from 400 plates, that span 100 years and 6 different plate-series. Comparing the RMS for the black (0.154 mag) and red (0.105 mag) set of points, the effect of correcting the irregular plate-variations gives about a 50\% improvement over radial-correction alone.
\\

{\em The eclipsing binary RY Cancri.} The lightcurve of RY Cnc contains 417 measurements of brightness obtained over 88 years. In figure~\ref{fig:RYcnc} one can see a scattering of faint points below the rest, which correspond to eclipses of the primary star by its fainter companion. The eclipses are very narrow as is typical for compact binaries. The lightcurve of RY Cancri is show folded in figure~\ref{fig:RYcncFold}, at its known binary period of 1.092943 days, to show the eclipse. The magnitudes plotted are corrected for spatial variations in the different plates (the red points in previous plot), and the symbols are as described above. The light-curve is plotted over two cycles, the first cycle shows estimated error bars for each point.
\\

{\em U Cancri, A semi-regular variable of the Mira type.} Figures~\ref{fig:UCnc} \& ~\ref{fig:UCncFold} provide an example of a red-giant star that is slowly pulsating. The star brightens and fades by over 4 magnitudes roughly every 300 days. During its minima U Cancri is not visible on some of our  plates, as can be seen from the upper limits (arrow symbols). Because the star's pulsations are not perfectly regular, when we fold the lightcurve the points from different cycles don't line up exactly. At least 2 distinct rising portions are visible in figure~\ref{fig:UCncFold}  beginning at phase 0.5. We were able measure a period of 304 days from the DASCH light-curve using the Lomb-Scargle periodogram.
\\

{\em U Geminorum, a CV.} U Gem is a famous cataclysmc variable (CV) that shows regular dwarf nova outbursts.  Outbursts, (due to an instability in the accretion disk) produce a 2-5  mag brightening, lasting for several days.  In the $\sim$22 year lightcurve-segment  shown in Figure~\ref{fig:UGem},  we see the quiescent luminosity (B$\sim$14) interspersed by dwarf-nova outbursts, each few years apart. Others occurred but were missed by the gaps in plate coverage, as verified by comparison with AAVSO records. An in depth analysis of this and other CVs appears in Paper III.
\\

{\em CQ Cancri - an RR-Lyra variable.} RR Lyrae stars are hot white giants pulsating with a constant period. The period and luminosity obey a relation enabling RR-Lyraes to be used to measure distances.The century-long DASCH  lightcurve is shown in Figure~\ref{fig:CQCnc}. Folding at at 0.524 days, which is the pulsation period of the star, reveals the characteristic RR Lyrae profile Figure~\ref{fig:CQCncFold}.

Further analysis of the DASCH light-curves of variable stars in M44 cluster is presented in Paper III.

\section{Acknowledgements}
We thank our colleagues at Harvard and beyond who have offered advice and expertise in photographic astronomy, digitization and software development.
In particular we thank A. Doane, G. Champine, and A. Sliski for their considerable contributions. DASCH has been supported by NSF grant AST-0407380 for which we are grateful.


\begin{deluxetable}{ccccccccc}l,l
\tabletypesize{\scriptsize}
\tablecaption{Plate Series used in this work.  \label{tab:plateseries}}
\tablewidth{0pt}
\tablehead{ \colhead{Series} &  \colhead{Location}  & \colhead{Size} &  \colhead{Sky Coverage}  & \colhead{Plate Scale} & \colhead{Image Scale} & \colhead{Color} & \colhead{Lens Aperture} & \colhead{Dates} \\
\colhead{Series} &  \colhead{(hemisphere)}  & \colhead{(inches)} &  \colhead{(\degr)}  & \colhead{(''/mm)} & \colhead{''/pixel @ 11$\mu$} & \colhead{} & \colhead{(inches)} & \colhead{}     }
\startdata
MC 	&  N  & 8x10  & 5x7       & 97.88      & 1.08    & B & 16 & 1909-1988  \\
RB	&  S  & 8x10  & 22x28  & 395.78    &  4.35    & B &  3  &  1928-1963  \\
RH	&  N  & 8x10  & 22x28  & 391.59   &  4.31      & B &  3  & 1928-1963  \\
BM	&  N  & 8x10  & 10x13  & 384.52   &  4.23      & B &  8  & 1882-1954  \\
I	&  N  & 8x10  & 9x11    & 163.23   & 1.80      &B &  8   & 1889-1946  \\
MA	&  N  & 8x10  &	5x7      & 93.68      &  1.03      & B  & 16 &  1905-1983   \\
MB   &   N  & 8x10  &	 11       & 389.99    &  4.29    & B  &  4  &  1914-1923  \\
\enddata
\tablecomments{{Further details of the Harvard Plate Collection may be found on the web at http://hea-www.harvard.edu/DASCH }}
\end{deluxetable}


\begin{figure}
\begin{center}
\includegraphics[angle=0,width=12cm]{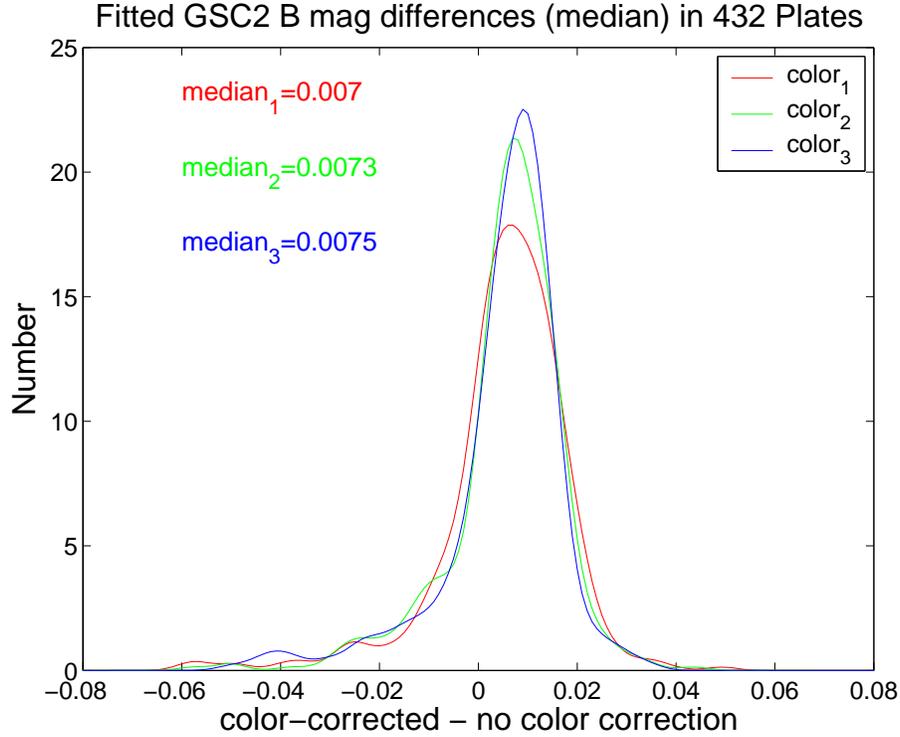}
\caption{{\bf Magnitude Errors due to Unsolved Color-Term.} If color-term corrections are not applied, there will be systematic shifts in the magnitude measured for the same star on different plates. Here we show the distribution of these errors for blue-sensitive plates under three scenarios for the color-term $C$: 1. Plate-dependent, 2. Series-dependent, 3. Constant=0.1. It appears that differences in color-response are not a significant source of scatter in lightcurves.   }
\label{fig:comparecolorterms}
\end{center}
\end{figure}

\begin{figure}
\begin{center}
\includegraphics[angle=0,width=14cm]{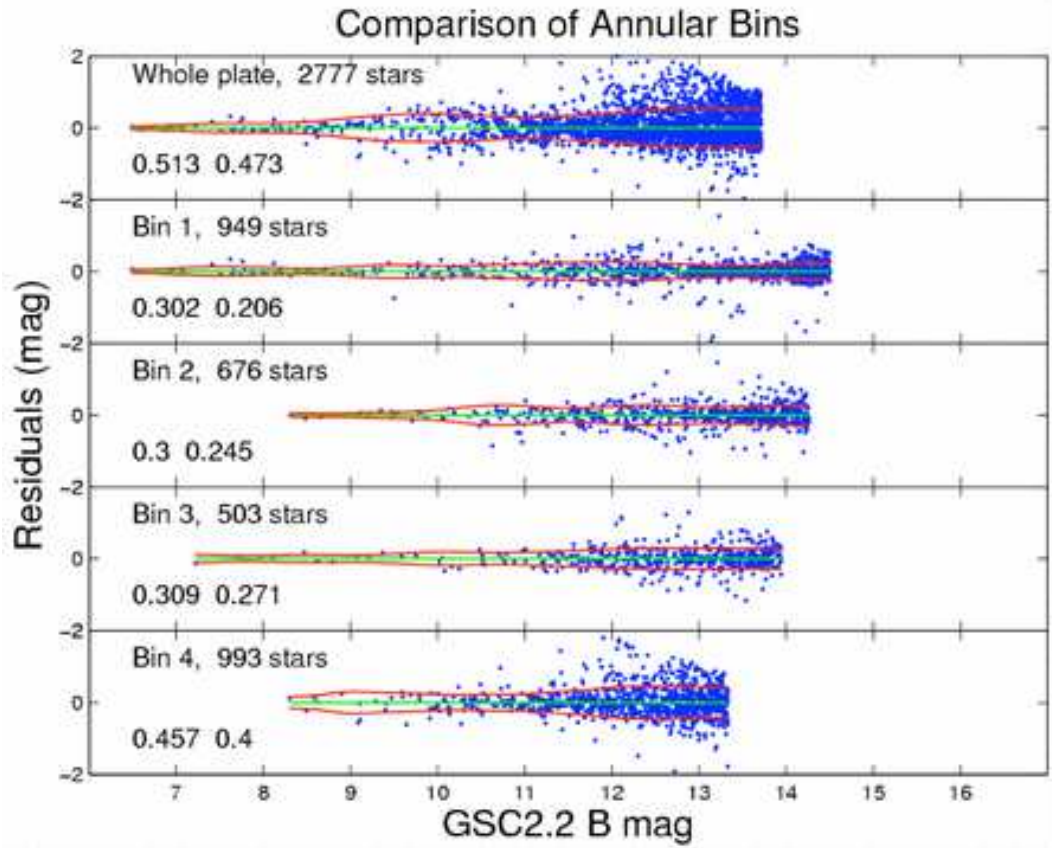}
\caption{{\bf Improvement in Photometric Accuracy with Radial Correction.} By fitting the calibration curve in a series of concentric annular regions on each plate, radially dependent effects (primarily vignetting) are greatly reduced. The calibration curves shown are four concentric annular equal-area bins. For comparison the top panel 
shows the calibration curve for the whole plate if radial effects are ignored. The improvement affected by the annular approach is evident from the visible scatter and  RMS values. }
\label{fig:spatialbins_rms}
\end{center}
\end{figure}

\begin{figure}
\begin{center}
\includegraphics[angle=0,width=12cm]{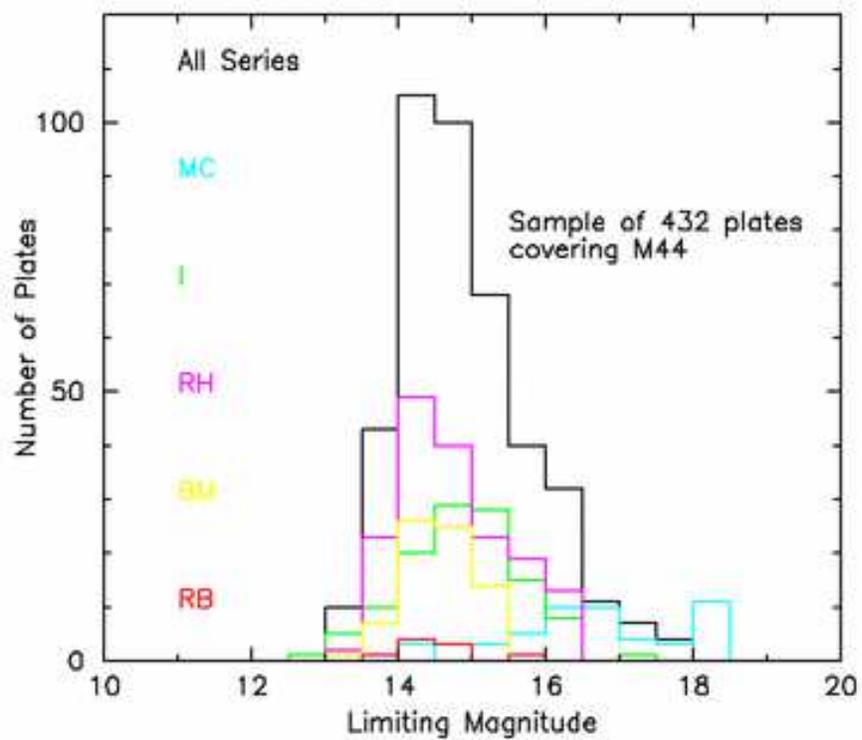}
\caption{{\bf Limiting Magnitudes (2$\sigma$) by Plate Series}}
\label{fig:limiting_mag_hist}
\end{center}
\end{figure}

\clearpage

\end{document}